\documentclass[num-refs]{wiley-article}
\usepackage{siunitx}
\usepackage{multirow}
\usepackage{amssymb}
\usepackage{hyperref}
\usepackage{subcaption}
\newcommand{\checkedbox}{$\rlap{$\large\checkmark$}\square$}

\papertype{Original Article}
\title{Tractography and machine learning: Current state and open challenges}

\author[1]{Philippe Poulin}
\author[2]{Daniel Jörgens}
\author[1]{Pierre-Marc Jodoin*}
\author[1]{Maxime Descoteaux*}

\affil[1]{Department of Computer Science, Université de Sherbrooke, Sherbrooke, Québec, Canada}
\affil[2]{Department of Biomedical Engineering and Health Systems, KTH Royal Institute of Technology, Stockholm, Sweden}
\affil[*]{Co-last author}

\corraddress{Philippe Poulin, Department of Computer Science, Université de Sherbrooke, Sherbrooke, Québec, Canada}
\corremail{Philippe.Poulin2@Usherbrooke.ca}

\fundinginfo{FRQNT; Université de Sherbrooke Institutional Chair in Neuroinformatics; NSERC Discovery grant from Pr Descoteaux and Jodoin}

\runningauthor{Philippe Poulin et al.}

\begin{document}

\maketitle

\begin{abstract}
Supervised machine learning (ML) algorithms have recently been proposed as an alternative to traditional tractography methods in order to address some of their weaknesses.  They can be path-based and local-model-free, and easily incorporate anatomical priors to make contextual and non-local decisions that should help the tracking process. ML-based techniques have thus shown promising reconstructions of larger spatial extent of existing white matter bundles, promising reconstructions of less false positives, and promising  robustness to known position and shape biases of current tractography techniques. 
But as of today, none of these ML-based methods have shown conclusive performances or have been adopted as a \textit{de facto} solution to tractography. One reason for this might be the lack of well-defined and extensive frameworks to train, evaluate, and compare these methods. 

In this paper, we describe several datasets and evaluation tools that contain useful features for ML algorithms, along with the various methods proposed in the recent years.
We then discuss the strategies that are used to evaluate and compare those methods, as well as their shortcomings.
Finally, we describe the particular needs of ML tractography methods and discuss tangible solutions for future works.

% Please include a maximum of seven keywords
\end{abstract}

\section{Introduction}

In the field of diffusion magnetic resonance imaging (dMRI), tractography refers to the process of inferring streamline structures that are locally aligned with the underlying white matter (WM) dMRI measurements~\cite{Jeurissen2017}.  
A simple approach to obtain such streamlines is an iterative process in which, starting from a seed point, an estimate of the local tissue orientation is determined and followed for a certain step length before repeating the orientation estimation at the new position.  
The tracking procedure may be deterministic~\cite{yeh2013deterministic,basser2000vivo} (at each point, the algorithm follows the strongest orientation) or probabilistic~\cite{behrens2007probabilistic,tournier2010improved,tournier2012mrtrix} (at each point, the algorithm samples a direction closely aligned with the strongest orientation).  
Tracking may also be  global as some methods recover streamlines all at once~\cite{reisert2011global,mangin2013toward,jbabdi2007bayesian}.  In between the local and  global methods is the category of shortest‐path methods, including front evolution, simulated diffusion, geodesic, and graph‐based approaches~\cite{Jeurissen2017}.
Ultimately, the collection of all trajectories created in that way is called a tractogram. 

In traditional methods, the estimate of the local tissue fiber orientation is usually inferred from an explicit and local model which fits the (local) diffusion data. 
These local models include diffusion tensor models~\cite{basser2000vivo,pierpaoli1996diffusion}, multi-tensor models~\cite{caan2010estimation}, and other methods that aim at reconstructing the fiber orientation distribution function (fODF) like constrained spherical deconvolution (CSD)~\cite{tournier2004direct,Descoteaux2009}, to name a few. 
However, the choice of the best model is by itself difficult \cite{schilling2018challenges,maier2017challenge}, as it depends on various factors such as data acquisition protocol or targeted WM regions, and therefore has a direct influence on the quality of an obtained tractogram~\cite{cote2013tractometer}.
Moreover, traditional methods based on local orientation alone are prone to make common mistakes, such as missing the full spatial extent of bundles and producing a great amount of false positive connections~\cite{maier2017challenge}.

Another important factor for the performance of a tractography method are the actual rules that regard the progression of a single step as well as simple global properties of an individual streamline. Traditional methods may define several engineered, or ``manually-defined'', high-level rules with the aim of improving the anatomical plausibility of the recovered tractogram. Instances of these are constraints on streamline length (i.e. filtering streamlines that are too long or too short), streamline shape (e.g. filtering streamlines with sharp turns), or progression rules that make streamlines ``bounce off'' the WM border when they are about to leave the WM mask with a certain angle~\cite{girard2014towards,Neher2017}. In the same way as modeling noise and artifacts, and defining the right local model, also the design of these high-level rules has a direct impact on the performance of a tractography method~\cite{cote2013tractometer,maier2017challenge}.

To address these inherent difficulties, recent proposals suggest that machine learning (ML) algorithms, supervised or unsupervised, may be used to implicitly learn a local, global or contextual fiber orientation model as well as the tracking procedure.
Approaches ranging from the application of self-organizing maps (SOM) \cite{duru2007som,duru2013self}, random forests (RF) \cite{Neher2017,Neher2015}, Multilayer Perceptrons (MLP) \cite{jorgens2018learning,Wegmayr2018}, Gated Recurrent Units (GRU) \cite{Poulin2017,benou2018deeptract,poulin2018deeptracker}, as well as Convolutional Neural Networks (CNN) \cite{wasserthal2018tract} and Autoencoders~\cite{lucena2018thesis}, have been employed at the core of tractography to drive streamline progression. 
Apart from the differences in their underlying architecture, these ML methods differ substantially in aspects of the exact problem formulation, e.g. definition of the input data to the model, modeling the predictions as a regression~\cite{Poulin2017,lucena2018thesis} or classification problem~\cite{Neher2015,jorgens2018learning}, or even the general tractography approach, i.e. whole-brain~\cite{Poulin2017,benou2018deeptract} or bundle-specific~\cite{wasserthal2018tract,lucena2018thesis,reisert2018hamlet,poulin2018deeptracker}.  The fact alone that these approaches differ in several aspects, makes it difficult to draw conclusions on the value of each of the individual modeling choices.

Furthermore, while the above mentioned approaches constitute the main ideas for applying ML directly to the process of tractography, machine learning and especially deep learning (DL) methods have been applied in related fields. 
Stacked U-Nets were proposed to segment the volume of individual white matter bundles from images of fODF peaks~\cite{wasserthal2018tractseg}. 
It was also suggested to predict fiber orientations from raw diffusion data based on convolutional neural networks (CNN)~\cite{Koppers2016}. 
Several ideas for streamline clustering or streamline segmentation have been proposed, including a CNN based on landmark distances~\cite{NgattaiLam2018}, a long short-term memory (LSTM)-based siamese network for rotation invariant streamline segmentation~\cite{Patil2017}, and a CNN approach for streamline clustering based on the sequence of their coordinates~\cite{Gupta2017, Gupta2018}.
Even though the mentioned works are closely related to tractography and contribute to the common goal of improved analysis of the white matter anatomy of the human brain, we restrict our focus exclusively on the direct application of ML (and especially DL) for tractography, with the explicit goal of producing streamlines and addressing the weaknesses of traditional methods. For that reason, we refer the interested reader to the respective references for more details.

An important factor for effectively advancing this field of research is a common and appropriate methodology for training and evaluating the performance of different approaches, which is currently lacking. Over the years, multiple challenges have been proposed to assess the performance of conventional tractography methods, and a clear and exhaustive review is provided by \citet{schilling2018challenges}.  However, we argue that the design of these challenges is typically inappropriate for ML methods.  In fact, the \textit{2015 ISMRM Tractography Challenge}~\cite{maier2017challenge} (along with the \textit{Tractometer} evaluation tool~\cite{cote2013tractometer}) has been adopted as the tool of choice for benchmarking new ML tractography pipelines~\cite{Poulin2017,Neher2017,Wegmayr2018,benou2018deeptract}.
Unfortunately, several inherent flaws arising specifically in the context of ML make it difficult to perform a fair comparison between the results obtained from different ML pipelines. In particular, diffusion data preprocessing is left to participants (\textbf{dissimilar inputs}), tracking seeds and a tracking mask are not always given (\textbf{varying test environment}), the test diffusion volume is sometimes used for training (\textbf{data contamination}), training streamlines are not provided (\textbf{disparate training data}), and testing on a single synthetic subject means that any computed estimator of a model's performance is unreliable (\textbf{small sample size}). 
Against the background of a prospectively increasing number of ML-based approaches tackling the problem of tractography, a carefully designed evaluation framework that appropriately addresses the specific requirements of ML methods has the potential to support and facilitate research in this field in the upcoming years.

In this paper, we follow a threefold strategy.  
First, we introduce the currently available datasets and evaluation tools along with useful features and weaknesses regarding machine learning.
Then, we provide a comprehensive review of existing ML-based tractography approaches and derive a set of key concepts distinguishing them from each other.
Subsequently, we identify and discuss the strategies for evaluation of tractography pipelines and identify issues and limitations arising when applied to ML-based tractography methods.
We finally describe important features for an appropriate evaluation framework the community ought to adopt in the near future to better promote data-driven streamline tractography and point out the potential advantages for research in data-driven streamline tractography.

\section{Annotated datasets and evaluation tools}  

Over the years, many diffusion MRI datasets were produced and annotated, either as part of a challenge or research papers.  
In this section, we overview several datasets that have been used to train and/or validate supervised learning algorithms for tractography. 
Specifically, we selected datasets that offer both diffusion data and streamlines.
Selected datasets also needed to have either clearly defined evaluation metrics, or to be large enough (more than 50 subjects) to be considered as standalone training sets.
We include datasets that are either publicly available or simply mentioned in a research paper without a public release.

We excluded datasets or challenges focused on non-human anatomy (e.g. rat or macaque), where the ground truth is harder to define and results might be harder to generalize to human anatomy (for data-driven algorithms), like the \textit{2018 VOTEM Challenge}~\cite{thomas2014anatomical} (\url{my.vanderbilt.edu/votem/}).
Moreover, we left out datasets focused only on pathological cases like the \textit{2015 DTI Challenge}~\cite{pujol2015dti}, because we consider it too early for data-driven tractography algorithms, at least until more conclusive results on healthy subjects.
We also excluded tractography atlases when tracking was done on a single diffusion volume, usually averaged over multiple subjects (e.g. HCP842~\cite{yeh2018population}), because results tend to be overly smooth and unsuited for ML methods.
However, we include a recent case when tracking was done for each subject: the 100-subjects WM atlas of \citet{zhang2018anatomically}.

While all the selected datasets are useful in one way or another for data-driven methods, they differ in multiple ways, which are detailed in the following subsections and summarized in Table \ref{tab:datasets}. 
The listed properties are the following: \vspace{-0.7cm}

\begin{itemize}
\item \textbf{Name}: The dataset name and reference
\item \textbf{Year}: The year of publication of the dataset or paper using the dataset
\item \textbf{Public}: Is the dataset (diffusion data and streamlines) publicly available?
\item \textbf{Real}: Is the diffusion data a real acquisition or is it simulated?
\item \textbf{Human}: Does the diffusion data represent the human brain anatomy?
\item \textbf{Subjects}: The number of subjects or acquisitions
\item \textbf{Bundles}: The number of bundles or tracks (if streamlines are available)
\item \textbf{GT}: Is a ground truth known?  For real acquisitions, streamlines validated by a human expert (e.g. neuroanatomist) are considered as GT despite the fact that these annotations are subject to inter-rater and intra-rater variations.
\item \textbf{Metrics}: Well-defined evaluation metrics are available with this dataset.
\item \textbf{Split}: Is the dataset split into a training and testing set that future works can rely on?
\end{itemize}
\vspace{-0.4cm}

Note that the notion of "ground truth" refers to an indisputable biologically-validated label assigned to an observed variable.  In medical imaging, such ground truth may be obtained with a biopsy~\cite{Thon17}, throughout careful complementary analysis~\cite{clevelandclinic} or by having several experts agreeing on a given diagnostic~\cite{Bernard16}.  Unfortunately, such restrictive definition of a ground truth is unreachable most of the time, especially for white matter tracks obtained from tractography, where no expert can truly assess the existence (or non-existence) of a given streamline in a human brain from MRI images only.  In fact, only synthetically-generated streamlines or man-made phantoms can be considered as real "ground truth".  Despite that, for the purpose of this paper, we also use the term "ground truth" for any data that has been manually validated by a human expert, typically a neuro-anatomist. In the medical imaging field, this annotated data would be called a \textit{gold standard}, while in the artificial intelligence community, it might be called \textit{weakly annotated data}.  Although such annotations do not meet the fundamental definition of a ground truth, it is nonetheless widely accepted by the medical imaging AI community~\cite{Menze14}.

\begin{table}[bt]
\caption{Annotated datasets.}
\label{tab:datasets}
\begin{threeparttable}
\begin{tabular}{r|c|c|cc|cc|c|cc}
\headrow
  \textbf{Name}                                   & \textbf{Year} & \textbf{Public}   & \textbf{Real}   & \textbf{Human}    & \textbf{Subjects}  & \textbf{Bundles}	& \textbf{GT}   & \textbf{Metrics}  & \textbf{Split} \\
  Fibercup \cite{fillard2011quantitative}                 & 2009          & \checkedbox       & \checkedbox     &                   & 1                  & 7        & \checkedbox   & \checkedbox       &  \\
  Simulated Fibercup \cite{wilkins2012development}        & 2012          & \checkedbox       &                 &                   & 1                  & 7        & \checkedbox   & \checkedbox       &  \\
  Tracula \cite{yendiki2011automated}              & 2011          &                   & \checkedbox     & \checkedbox       & 67                & 18        &  \checkedbox  &                   &  \\
HARDI 2012 \cite{daducci2014quantitative}               & 2012          & \checkedbox       &                 &                   & 2                  & 7        & \checkedbox   & \checkedbox       & \checkedbox \\
  HARDI 2013 \cite{daducci2013hardi}                      & 2013          & \checkedbox       &                 &                   & 2                  & 20       & \checkedbox   & \checkedbox       & \checkedbox \\
  ISMRM 2015 \cite{maier2017challenge,cote2013tractometer}& 2015          & \checkedbox       &                 & \checkedbox       & 1                  & 25       & \checkedbox   & \checkedbox       &  \\
  HAMLET \cite{reisert2018hamlet}                    	  & 2018          &        			  & \checkedbox     & \checkedbox       & 83                 & 12       &               &                   &  \checkedbox \\
  PyT (BIL\&GIN) \cite{chenot2018population}              & 2018          &                   & \checkedbox     & \checkedbox       & 410                & 2        &  \checkedbox  &                   &  \\
  BST (BIL\&GIN) \cite{rheault2018bundle}                 & 2018          &                   & \checkedbox     & \checkedbox       & 39                 & 5        & \checkedbox              & \checkedbox       &  \\
  TractSeg (HCP) \cite{wasserthal2018tractseg}                    & 2018          & \checkedbox       & \checkedbox     & \checkedbox       & 105                & 72       & \checkedbox              &                   &  \\
  \citeauthor{zhang2018anatomically} (HCP) \cite{zhang2018anatomically}                    	  & 2018          &        			  & \checkedbox     & \checkedbox       & 100                 & 58 + 198       & \checkedbox              &                   &   \\
\hline
\end{tabular}
\end{threeparttable}
\end{table}

\subsection{The FiberCup dataset and the Tractometer tool}

\paragraph{Original FiberCup Tractography Contest (2009)}

\citeauthor{fillard2011quantitative} proposed the \textit{FiberCup Tractography Contest}~\cite{fillard2011quantitative,poupon2010diffusion} in conjunction with the 2009 MICCAI conference. The goal was to quantitatively compare tractography methods and algorithms using a clear and reproducible methodology. They built a realistic diffusion MR 7-bundle phantom with varying configurations (crossing, kissing, splitting, bending).  
The organizers acquired diffusion images with b-values of 2000, 4000, and 6000~$\text{s/mm}^2$, and used isotropic resolutions of 3mm and 6mm, resulting in 6 different diffusion datasets.  Contestants were provided all datasets (but not the ground truth) and were free to apply any preprocessing they wanted on the diffusion images.  
Evaluation was done by choosing 16 specific voxels, or seed points, in which a unique fiber bundle is expected.  Participants were expected to submit a single fiber bundle for each of those seed voxels. Quantitative evaluation was done by comparing the 16 pairs of candidate and ground truth fibers using a symmetric Root Mean Square Error (sRMSE).

While the \textit{FiberCup Tractography Contest} makes a good test case for simple configurations, it does not represent a true human anatomy and does not impose a choice of b-value and preprocessing, which can induce significant differences in data-driven methods.  
Also, it does not provide any training streamlines, and is thus useful only as a validation tool for ML-based methods.  
Furthermore, the fact that it contains only one subject makes it hard to evaluate the true generalization capability of an ML method trained and tested on that dataset.  
However, it is the only dataset that provides seed points in order to have a uniform test environment, which is of utmost importance when comparing ML-based algorithms. 
In the end, it is unclear if for ML-based methods there would be any correlation  between a good performance on the FiberCup contest and good performance on human anatomy.

\paragraph{Tractometer evaluation tool (2013)}
In 2013, \citeauthor{cote2013tractometer} developed the \textit{Tractometer} evaluation tool, to be used alongside the original FiberCup data, with the aim of providing quantitative measures that better reflect brain connectivity studies.  
Using a Region of Interest (ROI)-based filtering method, a complete tractogram can be evaluated on global connectivity metrics, such as the number of valid and invalid bundles.
Furthermore, they propose two seeding masks: a complete mask (mimicking a brain WM mask), and a ROI mask (mimicking GM-WM interfaces).  
The tractometer was designed to address the fact that ``metrics are too local and vulnerable to the seeds given, and, as a result, do not capture the global \textit{connectivity} behavior of the fiber tracking algorithm''\cite{cote2013tractometer}.

\paragraph{Simulated FiberCup (2014)}

In 2014, \citeauthor{neher2014fiberfox} proposed a simulated version of the FiberCup, allowing new tracking algorithms to be tested using multiple acquisition parameters~\cite{neher2014fiberfox}. 
The simulated data can be used alongside the \textit{Tractometer} tool designed for the original phantom. 

\citeauthor{wilkins2012development} also developed a synthetic version of the FiberCup dataset, but did not publicly release the data~\cite{wilkins2012development}.  
Unfortunately, with regards to ML methods, the simulated FiberCup dataset suffers from the same shortcomings as the original FiberCup dataset as it contains only one non-human subject whose data is not split \textit{a priori} into a training and testing set.

\subsection{Tracula (2011)}
\citet{yendiki2011automated} published the Tracula method for automated probabilistic reconstruction of 18 major WM pathways.
It uses prior information on the anatomy of bundles from a set of training subjects.
The training set was built from 34 schizophrenia patients and 33 healthy controls, using a 1.5T Siemens scanner as part of a multi-site MIND Clinical Imaging Consortium~\cite{white2009global}.
The diffusion images include 60 gradient directions acquired with a b-value of 700~$\text{s/mm}^2$, along with 10 b=0 images, with an isotropic resolution of 2mm.
Whole-brain deterministic tracking was performed, followed by expert manual labeling using ROIs for 18 major WM bundles.
The dataset also includes a measure of the inter-rater and intra-rater variability for the left and right uncinate.

To our knowledge, this is the earliest apparition of a large-scale human dataset with expert annotation of streamlines. It is also the only dataset that includes a measure of inter-rater and intra-rater variability, which is a desirable feature for ML methods (also discussed later in Section~\ref{sec:guidelines}). Unfortunately, the complete set of diffusion images and streamlines has been incorporated into the method and is not public.

\subsection{HARDI Reconstruction Challenges}

\paragraph{HARDI Reconstruction Challenge (2012)}
 
\citeauthor{daducci2013hardi} organized the \textit{2012 HARDI Reconstruction Challenge}~\cite{daducci2013hardi} at the ISBI 2012 conference. The goal of the challenge was to quantitatively assess the quality of intra-voxel reconstructions by measuring the predicted number of fiber populations and the angular accuracy of the predicted orientations. 
A training set was released prior to the challenge, and a test set was used to score the algorithms.  As such, the 2012 HARDI dataset contains diffusion images but no streamlines.

Participants could request a custom acquisition (only once) by sending a list of sampling coordinates in q-space, and the organizers would then produce a simulated signal for the given parameters.  A $16\times 16\times 5$ volume was then produced, containing seven different bundles attempting to recreate realistic 3-D configurations.  The metrics proposed by the authors are ill-posed for ML-based methods because of the limited context available and the focus on local performances.  Like the \textit{FiberCup}, it would only be useful as a validation tool given the lack of training streamlines, a limited number of bundles (only seven) and a limited number of non-human subjects (only two).

\paragraph{HARDI Reconstruction challenge (2013)}

The \textit{2013 HARDI Reconstruction Challenge}~\cite{daducci2014quantitative} was organized one year later at the ISBI 2013 conference.  For ML-based methods, three improvements are relevant compared to the 2012 challenge: a more realistic simulation of the diffusion signal, a new evaluation system based on connectivity analyses and a larger set of 20 bundles.  
Indeed, data-driven methods try to learn an implicit representation without imposing a model on the signal, which means that the signal used for training and testing should be as close as possible to that in clinical practice.  Furthermore, the main benefit of data-driven methods is the ability to use context in order to make good predictions in a multitude of configurations, which means they have the potential to particularly improve connectivity analyses.  Therefore, it would be a better validation tool for ML-based methods than the \textit{2012 HARDI Reconstruction Challenge}.  Nonetheless, the dataset suffers from an inherent limitation as it contains only two non-human subjects.

\begin{figure}[tp]
    \centering
    \includegraphics[width=0.5\linewidth]{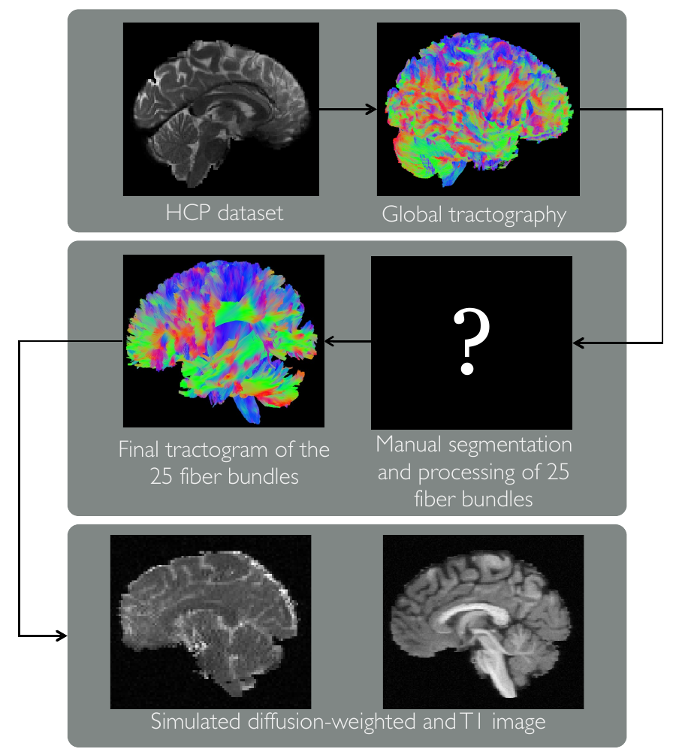}
    \caption{\textit{2015 ISMRM Tractography Challenge} data generation process (Taken from \url{www.tractometer.org})}
    \label{fig:ismrmchallenge}
\end{figure}

\subsection{ISMRM Tractography Challenge (2015)}
\label{sec:ismrmchallenge}
This dataset has been designed for a tractography challenge organized in conjunction with the 2015 ISMRM conference~\cite{maier2017challenge}.   During the challenge, participants were asked to reconstruct streamlines from a synthetic human-like diffusion-weighted MR dataset which was simulated with the aim of replicating a realistic, clinical-like acquisition, including noise and artifacts.  The available data consists of a diffusion dataset with 32 b=1000~$\text{s/mm}^2$ images and one b=0 image, with 2mm isotropic resolution, as well as a T1-like image with 1mm isotropic resolution.  Since all data was generated from an expert segmentation of 25 bundles, in theory, a perfect tracking algorithm should only produce exactly these specific bundles.  Unfortunately, as for the HARDI and FiberCup datasets, the \textit{2015 ISMRM Tractography Challenge} contains data from a limited number of subjects (only one) and lacks a clear separation between training and testing data.   Nonetheless, in combination with the \textit{Tractometer} tool \cite{cote2013tractometer}, this dataset has often been used to assess ML-based tractography methods.  Figure \ref{fig:ismrmchallenge} shows the data generation process for the challenge.

Once a tractogram has been generated using the challenge diffusion data, the tractometer tool uses a ``bundle recognition algorithm''~\cite{garyfallidis2018recognition} to cluster the streamlines into bundles.  The generated bundles are then compared to the ground truth, producing groups of ``valid bundles'' and ``invalid bundles'', depending on which regions of the brain the streamlines connect.  
Streamlines that do not correspond to a ground truth bundle are classified as ``No connections'' streamlines.   
The metrics computed by the modified \textit{Tractometer} for the \textit{Tractography Challenge} are as follows: \vspace{-0.3cm}
\begin{itemize}
\item \textbf{Valid bundles} (VB): The number of correctly reconstructed ground truth bundles.
\item \textbf{Invalid Bundles} (IB): The number of reconstructed bundles that do not match any ground truth bundles.
\item \textbf{Valid Connections} (VC): The ratio of streamlines in valid bundles over the total number of produced streamlines.
\item \textbf{Invalid Connections} (IC): The ratio of streamlines in invalid bundles over the total number of produced streamlines.
\item \textbf{No Connections} (NC): The ratio of streamlines that are either too short or do not connect two regions of the cortex over the total number of produced streamlines.
\item \textbf{Bundle Overlap} (OL): The ratio of ground truth voxels traversed by at least one streamline over the total number of ground truth voxels.
\item \textbf{Bundle Overreach} (OR): The ratio of voxels traversed by at least one streamline that do not belong to a ground truth voxel over the total number of ground truth voxels.
\item \textbf{F1-score} (F1): The harmonic mean of recall (OL) and precision (1-OR).
\end{itemize}\vspace{-0.3cm}

The definition of streamline-oriented metrics (VB, IB, VC, IC, NC) and volume-oriented metrics (OL, OR, F1) means that there is no single number that can fully assess the performance of an algorithm.  For example, deterministic methods often score higher on streamline-oriented metrics compared to probabilistic methods.  As such, a thorough review of all scores must be performed in order to properly compare algorithms, and in many cases, the choice of an algorithm over another may depend on a specific use-case (e.g. bundle reconstruction vs. connectivity analysis).

\subsection{HAMLET (2018)}
To validate their method, \citet{reisert2018hamlet} used a dataset of 83 human subjects  from two independent cohorts.  The first cohort comprises 55 healthy volunteers, all scanned by a Siemens 3T TIM PRISMA MRI scanner.  The second cohort has 28 volunteers scanned with a Siemens TIM TRIO.  The first cohort was used for training while the second one was used for testing.  Subjects in the second cohort were scanned twice for test-retest experiments, some unique characteristic to that dataset.  The reference streamlines were obtained by first tracking the whole brain with global tractography, and then by segmenting the streamlines for 12 bundles with a selection algorithm in MNI space.  Unfortunately, the recovered streamlines have not been manually validated by an expert.

\subsection{Datasets based on the BIL\&GIN database}

\paragraph{Bundle-Specific Tractography (2018)}

\citeauthor{rheault2018bundle} proposed a bundle-specific tracking method based on anatomical priors that improves tracking in the centrum semiovale crossing regions~\cite{rheault2018bundle}. Using multiple tractography algorithms, they tracked and segmented five bundles (Arcuate Fasciculus - AF left/right, Corpus Callosum - CC, Pyramidal Tracts - PyT left/right) in 39 subjects from the BIL\&GIN database~\cite{mazoyer2016bil}. To compare algorithms, they used an automatic bundle segmentation method based on clear anatomical definitions.  In addition, they defined several performance metrics, such as \textit{bundle volume}, \textit{ratio of valid streamlines}, and \textit{efficiency}. However, the tractograms and automatic bundle segmentation procedure were neither made public nor validated by an expert. 
Such a dataset, along with the evaluation procedure, could be extremely useful to assess if data-driven methods can reliably learn the structure of a specific bundle and reconstruct it in unseen subjects. 

\paragraph{A population-based atlas of the human pyramidal tract (2018)}
\citeauthor{chenot2018population} created a streamline dataset of the left and right PyT based on a population of 410 subjects~\cite{chenot2018population}, also from the BIL\&GIN database~\cite{mazoyer2016bil}. To do so, they combined manual ROIs along the bundles' pathway and the bundle-specific tractography algorithm of \citet{rheault2018bundle}.   
The quality of the segmentations and the high number of subjects would make this a noteworthy training dataset for data-driven methods.  Unfortunately for ML methods, only two bundles were examined. Furthermore, while the probability maps of the atlas have been rendered public, the tractograms are still unavailable.

\subsection{Datasets based on the HCP database}

\paragraph{TractSeg (2018)}

\citeauthor{wasserthal2018tractseg} proposed a data-driven method for fast WM tract segmentation without tractography~\cite{wasserthal2018tractseg}.  
In doing so, they built an impressive dataset of 72 manually-validated bundles for 105 subjects from the Human Connectome Project (HCP) diffusion database~\cite{van2013wu,glasser2013minimal}. 
Tractograms were obtained via a four-step semi-automatic approach:

\vspace{-0.3cm}
\begin{enumerate}
\item Tractography (Multi-Shell Multi-Tissue CSD~\cite{tournier2010improved})
\item Initial tract extraction (TractQuerier~\cite{wassermann2016white})
\item Tract refinement (Manual ROIs~\cite{stieltjes2013diffusion} + QuickBundles~\cite{garyfallidis2012quickbundles})
\item Manual quality control and cleanup
\end{enumerate}
\vspace{-0.3cm}

To the best of our knowledge, this is the largest public database to include both diffusion data and reference streamlines.  No further preprocessing of the diffusion data is needed because of the standard procedure of \cite{glasser2013minimal}.  
The authors defined volume-oriented metrics such as Dice score~\cite{taha2015metrics}, but did not offer any streamline-oriented metrics as their method predicts a volume segmentation.
The high number of subjects and bundles makes this a remarkable training set.

In a subsequent paper, the same authors re-used a subset of 20 bundles of the TractSeg dataset to train and validate their TOM ML algorithm~\cite{wasserthal2018tract}.  
However, as for original 72-bundle dataset, the TOM dataset does not come with a predefined set of training and testing data and no formal evaluation protocol that users could rely on has been proposed.

\paragraph{Zhang et al. (2018)}
\citet{zhang2018anatomically} built a WM fiber atlas using 100 HCP subjects.
They first generated streamlines for all subjects using a two-tensor unscented Kalman filter method~\cite{reddy2016joint}, and sampled 10,000 streamlines from each subject after a tractography registration step. Then, using a hierarchical clustering method, the authors generated an initial WM fiber atlas of 800 clusters. Finally, an expert neuroanatomist reviewed the annotations in order to accept or reject each cluster, and provided the correct annotations when the initial annotation was rejected.
The final, proposed atlas is comprised of 58 bundles (each composed of multiple clusters), along with ``198 short and medium range superficial fiber clusters organized into 16 categories according to the brain lobes they connect''~\cite{zhang2018anatomically}.

While the atlas is public, the sampled streamlines from the 100 subjects are all merged into the single template. In order for ML methods to benefit from this dataset, the streamlines would need to be separated back into the space of the particular original subjects. For this reason, we do not consider this dataset to be "public", in the context of machine learning.

\section{Machine learning methods for tractography}

For this review, we regard all supervised machine learning methods published in peer-reviewed journals, conferences or on arXiv (\url{arxiv.org}) and biorXiv (\url{biorxiv.org}).
We added the requirement that methods needed to be specifically designed for tractography, i.e. with the purpose of predicting a contextual streamline direction (and not reconstructing a local, non-conditional fODF or clustering streamlines).
This criterion includes whole-brain as well as bundle-specific tractography methods.
A summary of the main properties for all reviewed methods is provided in Table~\ref{tab:methods}.

\begin{table}[bt]
\caption{Main properties of data-driven methods for tractography.}
\label{tab:methods}
\begin{threeparttable}
\resizebox{\linewidth}{!}{
\begin{tabular}{cccccccc}
\headrow
\textbf{Method}                           	& \textbf{Model} 		& \begin{tabular}[c]{@{}c@{}}\textbf{Temporal}\vspace{-0.15cm}\\ \textbf{context}\end{tabular}      & \begin{tabular}[c]{@{}c@{}}\textbf{Spatial}\vspace{-0.15cm}\\ \textbf{context}\end{tabular}	& \begin{tabular}[c]{@{}c@{}}\textbf{dMRI}\vspace{-0.15cm}\\ \textbf{input}\end{tabular}	& \textbf{Prediction} & \begin{tabular}[c]{@{}c@{}}\textbf{Implicit}\vspace{-0.15cm}\\ \textbf{stop}\end{tabular} 	\\
\citeauthor{Neher2017}~\citep{Neher2017}            	& RF  				& 1 last direction     	& 50 samples	& Resampled DWI		& Classification	& \checkedbox \\
\citeauthor{Poulin2017}~\citep{Poulin2017}          	& GRU              	& Full                  & 1x1x1 voxel	& SH				& Regression		& \\
\citeauthor{poulin2018deeptracker}~\citep{poulin2018deeptracker}          	& GRU              	& Full                  & 1x1x1 voxel	& SH				& Regression		& \\
Benou et al.~\cite{benou2018deeptract} 					& GRU				& Full                  & 1x1x1 voxel   & Resampled DWI		& Classification	&  \checkedbox\\
\citeauthor{jorgens2018learning}~\citep{jorgens2018learning}  	& MLP   	& 2 last directions     & 1x1x1 voxel   & Raw DWI			& Regression		& \\
\citeauthor{Wegmayr2018}~\citep{Wegmayr2018}      		& MLP 				& 4 last directions     & 3x3x3 voxels  & SH				& Regression		& \\
\citeauthor{wasserthal2018tract}~\cite{wasserthal2018tract}	& CNN			& N/A                   & Entire WM     & fODF peaks 		& Regression		&  \\
\citeauthor{reisert2018hamlet}~\cite{reisert2018hamlet}    	& CNN-like     	& N/A                   & Entire WM     & SH				& Regression  		&  \checkedbox\\
\hline
\end{tabular}}
\begin{tablenotes}
\small
\item RF: Random Forest; MLP: Multilayer perceptron; GRU: Gated recurrent unit; CNN: Convolutional neural network; SH: Spherical harmonics coefficients; fODF: fiber Orientation Distribution Function; Implicit stop: indicates if a method learns its tracking stopping criterion or if it relies on a usual explicit criterion.
\end{tablenotes}
\end{threeparttable}
\end{table} 

\paragraph{Random Forest classifier}

To the best of our knowledge, \citeauthor{Neher2015} were the first to propose a machine learning algorithm for (deterministic) tractography~\citep{Neher2015}. 
They employ a RF classifier to learn a mapping from raw diffusion measurements to a directional proposal for streamline continuation. After collecting several of such proposals in a local neighborhood of the current streamline position (radius: 25\% of the smallest side length of a voxel), these are aggregated in a voting scheme to finally arrive at a single direction in which to grow the streamline.

To define reference streamlines for their experiments, the authors employ several tractography pipelines and train their classifier on each of the resulting tractograms. 
They determine the best trained model by evaluating the performance of each on a replication of the FiberCup phantom (based on the \textit{Tractometer} metrics of \cite{cote2013tractometer}). Finally, comparing the performance of the latter to all other reference pipelines, they report a superior performance of their tracking model over all other approaches. While tractograms were scored on a simulated phantom (i.e. no real anatomy), extended experiments presented in a subsequent paper~\cite{Neher2017} confirm the superiority of their approach on the \textit{2015 ISMRM Tractography Challenge} dataset (simulated data of a human anatomy).

\paragraph{Gated Recurrent Unit (GRU) Tracking}
Hypothesizing ``that there are high-order dependencies between'' the local orientation at a point of a streamline and the orientations at all other points on the same streamline, \citeauthor{Poulin2017} proposed a recurrent neural network (RNN)  based on a GRU \cite{Poulin2017} to learn the tracking process. 
Their method implements an implicit model mapping diffusion measurements to local streamline orientations which not only depends on measurements in a local context, but on all data previously seen along the extent of a particular streamline.  As opposed to~\cite{Neher2017,Neher2015}, the RNN model is implemented as a regression approach. 
In their experiments, the authors show that a recurrent model (when trained on reference streamlines obtained using deterministic CSD-based tractography~\cite{tournier2012mrtrix}) was able to outperform most of the original submissions in the \textit{2015 ISMRM Tractography Challenge} with respect to the \textit{Tractometer} scores (discussed in section~\ref{sec:ismrmchallenge}).

\paragraph{DeepTracker}
In a subsequent paper, \citet{poulin2018deeptracker} again suggested using a GRU, but in a bundle-specific fashion. While the model architecture is very similar, it was trained on a dataset of 37 real subjects, each with a curated set of streamlines for bundles. 
After training a single model for each of the selected bundles, the authors showed promising results compared to existing methods, perhaps indicative that the difficult task of learning to track streamlines necessitates more data than previously thought.

\paragraph{DeepTract}
More recently, \citet{benou2018deeptract} proposed a GRU-based recurrent neural network similar to that of \citet{Poulin2017}.  
In their method, they directly use the resampled diffusion signal as input to the model (like \citet{Neher2017}), in order to estimate a discrete, streamline-specific fODF representation which they refer to as ``conditional fODF'' (CfODF). Instead of predicting a 3D orientation vector using a regression approach, the authors implement their model as a classifier enabling them to interpret the probabilities obtained for discrete sampled directions (i.e. the classes) as the mentioned CfODF. This fODF-based formulation further allows for an inherently defined criterion for streamline termination based on the entropy of the CfODF. The proposed model can be employed for both deterministic and probabilistic tractography.

Like \citet{Poulin2017}, the authors trained and tested their method on the \textit{2015 ISMRM Tractography Challenge} dataset.  
They report results after training their method on the dataset ground truth as well as on streamlines obtained with the MITK diffusion tool~\cite{MITK}.

\paragraph{Multi-Layer Perceptron Point-Wise Prediction}
\citet{jorgens2018learning} propose a multi-layer perceptron (MLP) to predict the next step of a streamline. Like \cite{Neher2017,Neher2015,Poulin2017}, their method takes as input the diffusion signal and thus avoids explicit dMRI model-fitting.
The authors implemented different configurations of their proposed MLP such as three different input scenarios (point-wise input vs region-wise input with and without considering previous orientations), different approaches to aggregate the output (maximum likelihood, mathematical expectation of the categorical prediction and regression) as well as the voting scheme proposed by \citet{Neher2015}.  Results reveal that the best configurations are those having the previous two directions included in the input of the network thus showing that temporal context is a key component for data-driven tractography.  Also, the regression and classification approaches led to similar results and the use of region-wise information did not provide any substantial improvement over the use of point-wise information.

Like \citet{Poulin2017} and \citet{benou2018deeptract}, the authors trained and tested their method on the \textit{2015 ISMRM Tractography Challenge} dataset (but did not use the \textit{Tractometer} tool). Unfortunately, they did not estimate the tracking capabilities of their method as they only measured point-wise angular errors when predicting the next step of a streamline.

\paragraph{Multi-Layer Perceptron Regression Tracking}
A similar approach suggested by \cite{Wegmayr2018} employs a MLP to predict the next direction of a streamline through regression. At each point, the input of the model is given by all diffusion measurements in a cubic neighborhood, along with a certain number of previous steps for the current streamline.  In that way, the authors provide the ML model directly with diffusion information in a local neighborhood (spatial context) as well as a notion of ``history'' of the current streamline (temporal context).  
Defining their reference streamlines as tractograms obtained with a standard tractography method from \textit{in vivo} datasets, they train their model on three subjects from the HCP database.   
Experimental validations on the \textit{2015 ISMRM Tractography Challenge} dataset reveal that their model outperforms some ML methods  \cite{Neher2017, Poulin2017} in most \textit{Tractometer} metrics.   However, as demonstrated by low overlap scores, the authors acknowledge that their model produces ``rather confined bundles with little spread'', especially in contrast to \cite{Neher2017, Poulin2017}.  
While the strength of this model is to explicitly provide information from a local neighborhood, like for \citet{jorgens2018learning}, the notion of context along the streamline is limited and needs to be defined before training. Since the ideal temporal context (in terms of streamline length, or steps) is still unknown, this could potentially prohibit the model from taking advantage of all information relevant to streamline continuation.

\paragraph{Tract orientation mapping using an encoder-decoder CNN}
\citet{wasserthal2018tract} proposed a data-driven, bundle-specific tracking method.  
As opposed to the other ML methods reported in this paper, the authors do not try to directly reconstruct streamlines \textit{per se}. Instead, their proposed \textit{Tract Orientation Mapping} (TOM) method predicts bundle-specific fODF peaks that are then used by a deterministic tracking method.  First, CSD is used to extract three principal directions in all WM voxels.  Then, a U-Net CNN~\cite{Ronneberger15} is trained to map these fODF peaks to bundle-specific peaks, i.e. peaks that are only relevant for the streamlines of a given bundle. Their CNN takes as input 9 channels (the three fODF peaks) and outputs 60 channels, {i.e.} a 3D bundle-specific fODF vector for each of the 20 bundles they are looking to recover.  While the recovered bundle-specific peaks can be used in different ways, the authors show that using them directly as input to a deterministic MITK diffusion tractography gives some of the best results.
The approach was trained and tested on 105 HCP subjects, each with reference streamlines produced by a semi-automatic dissection of 20 large WM bundles (which they recently rendered public~\cite{wasserthal2018tractseg}).

\paragraph{HAMLET}
In a similar line of thought, in their HAMLET project (\textit{Hierarchical Harmonic Filters for Learning Tracts from Diffusion MRI}) \citet{reisert2018hamlet} map raw spherical harmonics of order 2 to a spherical tensor field.  In that sense, like \citet{wasserthal2018tract}, their ML method does not output streamlines but instead voxel-wise bundle-specific tensors that can subsequently be used as input to a classical tractography method. The magnitude of the produced $3\times 3$ tensor indicates the presence of a specific bundle whereas the tensor orientation predicts the local streamline direction.  Their method implements a multi-resolution CNN with rotation covariant convolution operations.  They trained and tested their method on two in-house datasets comprising a total of 83 human subjects.  The 12 bundles and their associated reference streamlines have been obtained with global tractography and automatic bundle selection method.  Unfortunately, the reference data was not manually validated by a human expert, and they did not perform any comparisons against other tractography methods.

\section{Results \& Discussion}
\subsection{Results on the 2015 ISMRM Tractography Challenge}

\begin{figure}[!t]
    \vspace{0.3cm}
    \centering
    \begin{subfigure}[b]{0.9\linewidth}
    	\centering 
            \includegraphics[width=\linewidth]{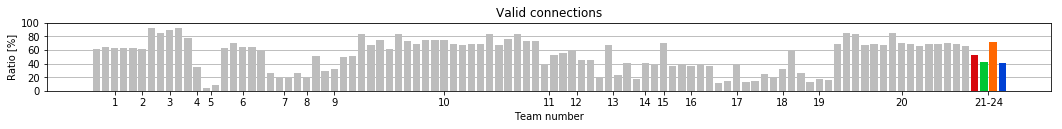}
        \caption{}
    \end{subfigure}\\
    \vspace{0.3cm}

    \begin{subfigure}[b]{0.95\linewidth}
    	\centering
        \begin{subfigure}[b]{.4\linewidth}
            \includegraphics[width=\linewidth]{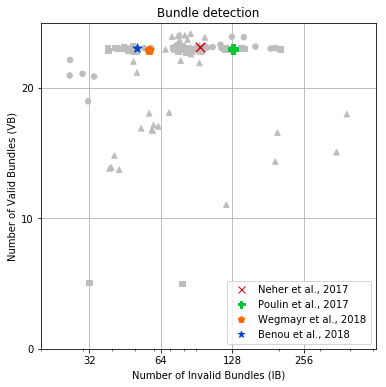}
            \caption{}
        \end{subfigure}
        \hspace{0.1cm}
        \begin{subfigure}[b]{.4\linewidth}
            \includegraphics[width=\linewidth]{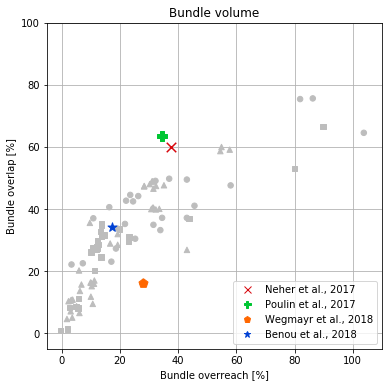}
            \caption{}
        \end{subfigure}
        
    \end{subfigure}
    \caption{\textit{2015 ISMRM Tractography Challenge} original submissions (1-20) and new results (21-24)}
    \label{fig:ismrmresults}
\end{figure}

The \textit{2015 ISMRM Tractography challenge} is the only dataset that has been used to assess performance of several data-driven tractography methods and is thus, as of today, the only available common ground on which to compare methods.  
It was used by four different papers namely, the Random-Forest of \citet{Neher2017}, the GRU of \citet{Poulin2017} and \citet{benou2018deeptract}, and the MLP of \citet{Wegmayr2018}.  
Experimental results reported by the authors have been transcribed in Table \ref{tab:tractometer}, and compared with original submissions in Figure~\ref{fig:ismrmresults}.  
Note that the metrics marked as \textit{not available} (N/A) are those the authors did not report in their original paper.

As can be seen, results vary a lot and there is no clear trend showing which method performs best, especially given the nature of the evaluation metrics. 
As mentioned in section \ref{sec:ismrmchallenge}, methods can be evaluated using both \textit{streamline-oriented} metrics and \textit{volume-oriented} metrics, which are not always correlated.  
For example, a method may have a large number of valid connections but a low overlap (like the MLP of \citeauthor{Wegmayr2018}) which means that although the model was able to recover most valid bundles, the generated streamlines do not properly cover the spatial extent of those bundles.
Also, a method can be more conservative and score best in terms of invalid connections and overreach like the GRU of \citeauthor{benou2018deeptract}, but at the same time have a low ratio of valid connections and a poor bundle overlap.
On the other hand, the Random-Forest of \citeauthor{Neher2017} does not score best in any category, but is competitive according to all metrics (its large F1-score underlines that it is a more balanced method compared to MLP and DeepTract).  
On top of that, all methods were trained and evaluated differently, so any comparison based on the reported results should be done with extreme care. 

\begin{table}[!t]
\caption{Tractometer results.  The Bundles and Connections (\%) metrics are {\em streamline-oriented metrics} whereas the Avg. bundle (\%) metrics are {\em volume-oriented metrics}}
\label{tab:tractometer}
\begin{threeparttable}
\begin{tabular}{c|rr|rrr|rrr}
\headrow
\textbf{Model}                  & \multicolumn{2}{c}{\textbf{Bundles}}                             & \multicolumn{3}{c}{\textbf{Connections (\%)}}                                                        & \multicolumn{3}{c}{\textbf{Avg. bundle (\%)}}                                                   \\
                                & \multicolumn{1}{c}{\emph{Valid}} & \multicolumn{1}{c}{\emph{Invalid}} & \multicolumn{1}{c}{\emph{Valid}} & \multicolumn{1}{c}{\emph{Invalid}} & \multicolumn{1}{c}{\emph{No connection}} & \multicolumn{1}{c}{\emph{Overlap}} & \multicolumn{1}{c}{\emph{Overreach}} & \multicolumn{1}{c}{\emph{F1-score}} \\
Random-Forest~\citep{Neher2017}            &               \textbf{23} &                          94 &                        52 &                         N/A &                               N/A &                          59 &                            37 &                       61 \\
GRU~\citep{Poulin2017}          &               \textbf{23} &                         130 &                        42 &                          46 &                       \textbf{13} &                 \textbf{64} &                            35 &              \textbf{65} \\
MLP~\citep{Wegmayr2018}         &               \textbf{23} &                          57 &               \textbf{72} &                         N/A &                               N/A &                          16 &                            28 &                       26 \\
GRU (DeepTract)~\citep{benou2018deeptract} &               \textbf{23} &                 \textbf{51} &                        41 &                 \textbf{33} &                                23 &                          34 &                   \textbf{17} &                       44 \\
\hline
\end{tabular}
\end{threeparttable}
\end{table}

\subsection{The 2015 ISMRM Tractography Challenge as an evaluation tool for ML algorithms}
\label{sec:shortcomings}

As mentioned before, the \textit{2015 ISMRM Tractography Challenge} has been adopted as the \textit{de facto} evaluation tool to compare ML tractography methods.  
However, the strengths and weaknesses of that tool should be thoroughly reviewed to understand and trust any technique reporting results with it.  
In this section, we present what we consider to be important issues with the way in which this tool has been used to assess the performance of data-driven methods.  
In particular, we detail the discrepancies between the four ML-based methods, differences that may explain some of the results in Table~\ref{tab:tractometer} and potentially undermine  any conclusion that one could draw from it.  
Let us mention that some of these issues with the 2015 ISMRM dataset are typical for the field of tractography as a whole.

Table \ref{tab:challenge_preprocess} presents a summary of the differences in how the tool is used.  
Note that the \textit{not available} (N/A) mark is used for any information the authors did not mention in their original paper.

\begin{table}[tp]
\caption{Differences in data}
\label{tab:challenge_preprocess}
\begin{threeparttable}
\begin{tabular}{cccccc}
\headrow
\textbf{Method}                     & \textbf{Preprocessing}           & \textbf{WM mask}  & \textbf{Training subjects}    & \textbf{Reference streamlines}   \\
Random-Forest \citep{Neher2017}     & \emph{dwidenoise} + \emph{dwipreproc}                   & Not needed        & 5 HCP subjects                        & CSD (Deterministic) \\
GRU \citep{Poulin2017}              & None                               & Ground Truth      & Challenge subject                     & CSD (Deterministic) \\
MLP \citep{Wegmayr2018}             & \emph{dwipreproc}                               & N/A               & 3 HCP subjects                         & iFOD (Probabilistic) \\
DeepTract \cite{benou2018deeptract} & N/A                           & Not needed        & Challenge subject                     & Q-Ball (Probabilistic)
\end{tabular}
\end{threeparttable}
\end{table}

\paragraph{Dissimilar inputs}
The four ML methods use a different preprocessing pipeline.  
Among the proposed algorithms, two applied MRtrix's \textit{dwidenoise} or \textit{dwipreproc} (\url{www.mrtrix.org}), another one denoised using \cite{manjon2013diffusion} and corrected for eddy currents and head motion, and another one did not apply any preprocessing at all.
Moreover, some used the diffusion signal directly as input, while others resampled it to a specific number of gradient directions. 
In some cases, spherical harmonics were fitted to the signal and the SH coefficients were fed as input to the model.
Finally, the non-recurrent models are also given a variable number of previous streamline directions as input.

The output of each of these pipelines contain various degrees of information.
For example, fODF peaks are in theory already aligned with the major WM pathways, and information may be lost depending on the specific model used to recover the peaks from the diffusion signal.
On the other hand, using the raw diffusion signal might contain more information but is more difficult to understand and process, and thus a data-driven model might require more capacity to use such an input.
Without a thorough investigation of the information contained in each output, any variations in the \textit{Tractometer} results could be attributed to the variations in preprocessing. Since we currently do not have any indication of what is useful for data-driven algorithms, it is impossible to compare ML methods if they do not use the same input data.

\paragraph{Varying test environment}
Since no white matter mask is provided, it must be computed by each participant in case it is needed for tracking.  Out of the four ML methods that were evaluated on the challenge, two needed WM masks; one used the ground truth mask, and the other did not mention how the mask was computed. Furthermore, since no tracking seeds are supplied with the data either, their arrangement entirely depends on the WM mask (and the number of seeds per voxel, which is also not given).

Given the nature of streamline tractography, small variations of the tracking mask or the tracking seeds could have a substantial impact on the resulting streamlines and by that also on the obtained evaluation metrics. Also, even though computing a stopping criterion within the algorithm is a worthy improvement, it is a different task than tracking, and should be evaluated separately. Consequently, all methods should be provided the same tracking mask and seeds to reduce as much as possible the number of free variables during evaluation.

\paragraph{Data contamination}
The use of ML methods requires special care when dealing with available data.   
Since machine learning models are obtained by deriving implicit rules \textbf{directly from given data} (i.e. \textit{training data}), testing the true generalization capabilities of these rules must be done using a \textbf{different and unseen set of data} (i.e. \textit{test data}).  

Two methods suffer from data contamination, or \textit{leakage} \cite{kaufman2012leakage}: the GRU in \cite{Poulin2017} and the MLP in \cite{Wegmayr2018}.  Here, data contamination refers to the usage of the same diffusion data for training and testing.  This means that the true generalization capabilities of the tested method on new, thus unseen subjects are still unknown, since the model has already seen the specific diffusion patterns that are needed in order to ``explore'' at test time, and therefore has been given an ``unfair'' advantage.

\paragraph{Disparate training data}
All methods used different reference streamlines and subjects for training. 
As mentioned earlier, some employed the test diffusion data directly, while others relied on a varying number of subjects from the HCP database. 
Two methods used deterministic CSD tracking~\cite{tournier2012mrtrix} to generate reference streamlines, one used QBI tracking~\cite{aganj2009odf} (probabilistic) and the last one used iFOD tracking~\cite{tournier2010improved} (also probabilistic).
In order to provide a uniform basis for comparison, the same comprehensive streamline training set should be available to every algorithm.

\paragraph{Simulation as a substitute for human acquisition}
While the diffusion signal of the 2015 ISMRM dataset is typical of that of a human brain, it is nonetheless obtained through simulation.  
As such, results on that dataset should not be seen as a measure of future performance on real human subjects, at least not without further empirical evaluation.  
Furthermore, at the given resolution and using this particular configuration of 25 bundles, false positive streamlines that would otherwise be plausible given the underlying anatomy of a real scan might be impossible to avoid. 
Indeed, some authors tried training their models using the ground truth bundles, and still produced over 50 invalid bundles in both cases~\cite{Neher2017,benou2018deeptract}.

\paragraph{Small sample size}

The \textit{2015 ISMRM Tractography Challenge} dataset has only one subject, which makes it hard to assess the future performance of a data-driven algorithm~\cite{raudys1991small}.
In order to compute unbiased estimates of future performance, a richer testset with more subjects is needed. Also, given more subjects, bootstrapping methods~\cite{efron1994introduction} (i.e. sampling with replacement) could help to build more accurate estimators.

\subsection{Other results}
Some authors report local performance measures, such as the mean angular error~\cite{jorgens2018learning}.  However, local metrics do not take into account compounding errors, which can have a major effect on global structure.  Consequently, global evaluation metrics should be preferred.

\begin{figure}[tp]
    \centering
    \includegraphics[width=0.7\linewidth]{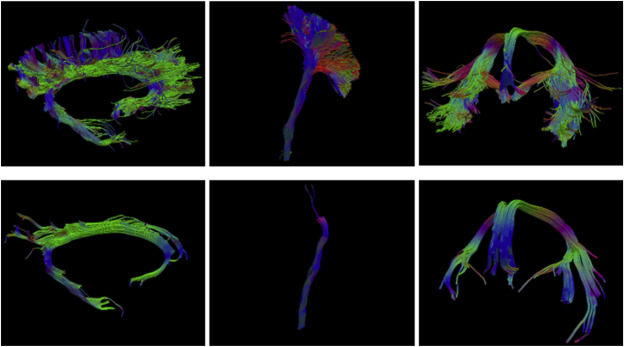}
    \caption{Comparison between the RF of \citeauthor{Neher2017} (top row), and classical deterministic CSD streamline tractography (bottom row). Results obtained on HCP subject 992774. \textit{(Taken from \cite{Neher2017} with authorization from the authors)}}
    \label{fig:neher_comparison}
\end{figure}

\begin{figure}[tp]
    \centering
    \includegraphics[width=0.95\linewidth]{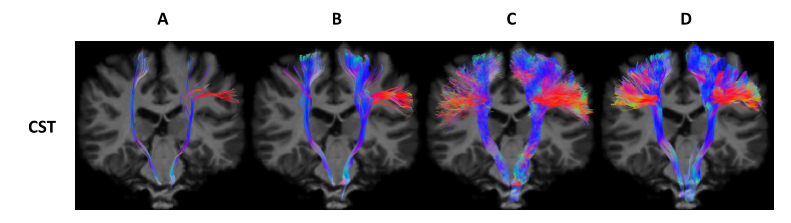}
    \caption{Comparison of various tracking methods: A: Deterministic, B: Deterministic Bundle-Specific (DET-BST)~\cite{rheault2018bundle}, C: Probabilistic particle filter BST (PROB-PF-BST)~\cite{girard2014towards}, D: DeepTracker~\cite{poulin2018deeptracker}. Results obtained on a BIL\&GIN subject. \textit{(Taken from \cite{poulin2018deeptracker} with authorization from the authors)}}
    \label{fig:deeptracker_comparison}
\end{figure}

Tractography papers often report a visual evaluation on unseen, \textit{in vivo} subjects, as a qualitative evaluation. For example, Figures~\ref{fig:neher_comparison} and~\ref{fig:deeptracker_comparison} compare some of the proposed data-driven approaches with standard tractography methods on white matter bundles with known anatomy. However, in absence of a ground truth or the expertise of a neuroanatomist, it is hard to draw definitive conclusions on the quality of such results.
In addition, \citet{reisert2018hamlet} presented correlation plots to assess reproducibility, but only offered qualitative comparisons with the reference streamlines without any quantitative results.  To gain trust in these data-driven methods, a more rigorous approach is needed.

Finally, most ML methods offer a reduction in computation time compared to traditional methods. This is a non-negligible benefit, should these methods be adopted in practice.

\subsection{Proposed guidelines for a data-driven tractography  evaluation framework}
\label{sec:guidelines}
Considering the ML tractography evaluation issues previously underlined, we discuss in this section the fundamental elements of a better framework we believe the community should adopt in the upcoming years.  
We start with the essential characteristics such a framework should have, followed with useful features.

\paragraph{Essential characteristics}
First and foremost, an ideal data-driven tractography evaluation framework should come with a public and free-to-use dataset that anyone could easily rely on.  
The dataset should include images of real human acquisitions along with a careful expert selection of ground truth streamlines.
It is important to avoid any bias towards a specific tractography algorithm. In order to achieve this, the streamlines could be first generated by a large number of different (and ideally orthogonal) deterministic, probabilistic and global algorithms and then segmented by expert annotators according to strict anatomical definitions for a given number of bundles.  While such manual annotation would be tedious, time consuming and even error prone, we consider this an indispensable step towards building a realistic and useful dataset for ML-based development. The need for such a gold standard that quantifies human variability is well-known in other fields, such as automatic image segmentation, cell counting or in machine learning \citep{kleesiek2016virtual,entis2012reliable,boccardi2015delphi,piccinini2014improving}.  
Despite the fact that simulated brain images come with a pixel-accurate set of ground truth streamlines that can be generated in a matter of seconds, by definition synthetic diffusion signals are over-simplistic pictures of real data and, as such, cannot provide any guarantee of subsequent performance for data-driven methods on real data. 

Although there is no consensus regarding the most desirable features a ML tractography algorithm should have and how it should be evaluated, by its very nature, any ML evaluation framework should aim at measuring how an algorithm can faithfully reproduce a task it was trained for. As such, a reasonable dataset should include a sufficiently large number of well-separated training and testing images.  Thus, statistics resulting from such a dataset would not suffer from contamination and the reported metrics would be reliable and unbiased estimates of the true generalization power of a ML algorithm.  In addition, to ensure that the observed differences between multiple algorithms are resulting from the intrinsic properties of the model and not caused by some feature of the evaluation framework, the number of free variables should be reduced to a minimum. 
Consequently, the tracking masks and seeds should be provided together with  clearly preprocessed diffusion data, so that the proposed methods can be evaluated in equal conditions. There should be multiple "classes" of input data, depending on whether an algorithm supports DWI samples, SH coefficients or fODF peaks.
Furthermore, the initial diffusion signal should have the same statistical properties for the training and the testing set.  Finally, the acquired images should ideally be acquired at different MRI scanners with different acquisition protocols in order to avoid overfitting issues.

Evaluation metrics should also be bound to the purpose of tractography algorithms.  
Considering that tractography is mostly used for bundle reconstruction, tractometry studies and connectivity analyses, an ideal evaluation framework should include two sets of metrics : 1) metrics measuring how a ML method can faithfully reproduce a set of predefined bundles it was trained to recover (tractometry), and 2) metrics measuring how it can connect matching regions of the brain, i.e. produce valid connections (connectivity).  Furthermore, since many applications use tractography algorithms to produce a large number of streamlines (with many false positives), which are then filtered out by a post-processing algorithm such as RecoBundles~\cite{garyfallidis2018recognition}, the framework should report results before and after post-processing. 
This would underline the true recall power of a data-driven algorithm, which is a fundamental characteristic of tract-based and connectivity-based applications~\cite{maier2017challenge}.

Lastly, the size of an ideal dataset is of primary importance.  
While a small-sized dataset could be prone to overfitting, it would be costly to create a very large dataset and also difficult to ensure a coherent manual annotation.  
One rule of thumb that can be used to identify the "correct" size of a dataset is through the inspection of the learning curve of several ML models~\cite{Beleites14}.  These curves show the model performance as a function of the training sample size.  Typically, the performance of several models saturates for a sufficient dataset size.  Although imperfect, this procedure is a good heuristic for estimating the size of the dataset.

\paragraph{Other useful features}
Despite any thorough manual annotation protocol, manually annotated bundles can be subject to non-negligible  inter-rater and intra-rater variability.  As such, a useful characteristic of a ML tractography dataset would be a measure of those variations.
This would be obtained by having several experts annotating the dataset, and at least one expert annotating it twice or more times.  Such measures would provide a minimal bound beyond which a data-driven algorithm could be considered "as good as an expert".  
Another very useful tool would be an openly accessible online evaluation system.  
Given such a system, people could upload their test results in order to compare them with the test ground truth.  In that way, an automatic ranking procedure similar to that of Kaggle could be used to sort various ML algorithms based on their achieved scores.  
While no ranking method is perfect, it would nonetheless provide a common evaluation framework that people could rely on.

An ideal dataset would also cover the whole field of diffusion MRI acquisition protocols, from HCP-like research acquisitions to clinical acquisitions. It would include single b-value as well as multiple b-values data, along with more sophisticated acquisition protocols such as b-tensor encoding.
It would also need low resolution images together with high-resolution images.
Since data harmonization is also a problem for data-driven algorithms, acquisition from several sites are needed for test-retest studies.
Annotated pathological cases would complete the dataset by allowing careful preliminary studies on how ML-based methods can be relied on in unhealthy patients.

Finally, since tractography is used more and more in pre-clinical applications, a subset of manually annotated rodent or macaque brains would be of great interest to train and test future ML algorithms (like the \textit{2018 VOTEM Challenge}~\cite{thomas2014anatomical}, for example).

This is, of course, the ultimate wish list. But, in the era of open data and open science, it needs to be done by the community, for the community. 
We can already see this work in progress with more and more accessible and reproducible data being published every year.

\section{Conclusion}
In this paper, we provided an exhaustive review of the current state of the art of machine learning methods in the field of tractography.  We described the existing datasets that comprise both diffusion data and reference streamlines, which could generally be useful for new tracking methods based on ML. In particular, we thoroughly examined the widely used evaluation tool for data-driven tracking methods, the \textit{2015 ISMRM Tractography Challenge}, and detailed flaws and shortcoming when used to assess data-driven algorithms. Based on our findings, we suggested good practices that we believe would foster the development of a new evaluation framework for ML-based tractography methods with the potential to effectively advance this field of research.

There is no doubt that machine learning tractography will have an important role to play in the future to solve some of the open problems of tractography. At the moment, however, all existing methods show theoretical potential and in limited test cases. Methods have yet to make solid demonstrations of their performance and efficiency in practice. There is still no ML-based tractography tool that is a scalable and usable on any given diffusion MRI dataset. This is true for healthy datasets but even more so for pathological brains. Hence, it is fair to say that ML-based tractography is still at its infancy and is not ready for "prime-time", but is nonetheless a very fertile field of research to make meaningful contributions to the field of connectivity mapping.

\bibliography{main}

\end{document}